\documentclass[aps,preprint]{revtex4}%
\usepackage{amsfonts}
\usepackage{amsmath}
\usepackage{amssymb}
\usepackage{graphicx}%
\begin{document}
\preprint{ }
\title[No tunneling limit for lasers]{Incompatibility of the tunneling limit with laser fields}
\author{H. R. Reiss}
\affiliation{Max Born Institute, 12489 Berlin, Germany}
\affiliation{American University, Washington, D.C. 20016-8058, USA}


\begin{abstract}
The\textit{ Schwinger limit} refers to longitudinal electric fields that are
sufficiently strong to \textquotedblleft polarize the vacuum\textquotedblright%
\ into electron-positron pairs by a tunneling mechanism. Laser fields are
transverse electromagnetic fields for which the Schwinger limit has no
relevance. Longitudinal and transverse fields are fundamentally different
because of the different values of the $F^{\mu\nu}F_{\mu\nu}$ Lorentz
invariant that characterizes the fields. One aspect of this difference is the
zero-frequency limit, that exists for longitudinal fields, but is ill-defined
for transverse fields. The goal of approaching the Schwinger limit with
sufficiently strong lasers is thus not a possibility. Tunneling transition
rates are characterized by an exponential behavior of the form $\exp\left(
-C/E\right)  ,$ where $E$ is the magnitude of the applied electric field and
$C$ is a system-dependent constant. Searches for such behavior within a
Coulomb-gauge treatment of laser-induced processes are shown to fail.

\end{abstract}
\date[31 January 2013]{}
\maketitle

Laser-induced ionization is widely regarded as a tunneling process.
Nevertheless, the tunneling model applies only to a limited range of field
parameters, and the use of this model outside that range leads to severe
failures of qualitative and quantitative predictions. Among these failures is
the prediction of a \textquotedblleft tunneling limit\textquotedblright\ for
laser-induced processes and the association of this limit with a simple
behavior as the field frequency approaches zero. An example is the concept of
a Schwinger limit\cite{sauter,schwinger} for laser-induced electron-positron
pair production in the vacuum. A single cause underlies these misconceptions:
the tunneling model and the associated tunneling limit and Schwinger limit
apply to longitudinal fields, whereas lasers produce the fundamentally
different transverse fields.

The demonstration of the above-stated conclusions follows two tracks. In one,
the theme is elaborated that the tunneling concept has application only to
longitudinal fields, whereas laser fields are transverse. The other approach
is to show that the characteristic exponential behavior for tunneling-type
processes does not arise from the Coulomb-gauge or the velocity-gauge (VG) approaches.

A simple qualitative argument leads to the above conclusions. The
transformation by an applied field of an impenetrable barrier to one that is
penetrable requires the representation of the applied field by a scalar
potential. In atomic ionization, the static Coulomb field responsible for the
binding of an electron in an atom has superimposed upon it the electric field
envisioned as an oscillating quasistatic electric (QSE) field. The fact that
the combined Coulombic and QSE fields can be represented as a superposition of
scalar fields is a prerequisite for graphic expression of the tunneling
phenomenon. The pictorial view of pair production has the $-mc^{2}$ upper
boundary for negative energy electrons and the $+mc^{2}$ lower boundary for
positive energy electrons so tilted by a QSE field as to allow tunneling from
the negative energy continuum to the positive energy continuum, which is a
pair production mechanism. With the presumed \textquotedblleft laser
field\textquotedblright\ represented by a scalar potential $\phi$, the
implication is that there is no vector potential: $\mathbf{A}=0$; and thus
there is no magnetic field: $\mathbf{B}=0.$ The absence of a magnetic field
means that the fundamental Lorentz invariant is positive: $\mathbf{E}%
^{2}-\mathbf{B}^{2}>0$ (Gaussian units). A positive value for this invariant
is the hallmark of a QSE field; that is, it is a longitudinal field. This is
in contrast to a transverse (plane-wave) field that is characterized by
$\mathbf{E}^{2}-\mathbf{B}^{2}=0$. Strong fields accentuate the mismatch of
the Lorentz invariant between longitudinal and transverse fields.

An appropriate value for the Lorentz invariant is related to the fact that
when there are no external sources or currents to sustain a field, as is the
case with a laser beam, the Maxwell equations can be satisfied only by a
transverse field. A transverse field is configured such that the electric
field vector, the magnetic field vector, and the propagation vector form a
mutually orthogonal triad. In contrast to this unique situation, a QSE field
(and its zero-frequency limit -- a constant electric field) requires external
sources to generate it, and the sole inherent direction in the process is the
direction of the electric field vector. For example, the field between the
plates of a capacitor is a constant electric field if there are fixed, unequal
charges on the plates, or it will be a QSE field if the capacitor is part of
an AC circuit. In both cases, an external source is required to sustain the field.

An essential distinction between transverse and longitudinal fields is that a
transverse field requires a magnetic component if it is to be a propagating
field. There is no such thing as a propagating longitudinal field. It can
oscillate with time, but it cannot propagate. (A seeming exception to this
statement comes from the introduction by Keldysh\cite{kel} of a putative
Volkov solution in a problem described within the length gauge (LG), where
only a scalar potential exists. However, the Volkov solution, and the gauge
transformation employed to transform from the Coulomb gauge to the length
gauge, are both dependent on a vector potential that does not exist in the
length gauge. A proposed solution to this dilemma is to replace the vector
potential by an integral of the electric field over all times prior to the
laboratory time. This causes the Volkov solution in the length gauge to be a
non-local quantity. This \textquotedblleft Volkov solution\textquotedblright%
\ in a gauge that does not have a true Volkov solution is treated in the
dipole approximation, where it does not possess the propagation property.)

A major distinction between longitudinal fields and transverse fields arises
from contrasting properties as the field frequency $\omega$ approaches zero.
When $\omega\rightarrow0$, a longitudinal field approaches a constant electric
field, where exact solutions are simple and well-known. This has led some
authors to base the soundness of tunneling-type theories on a presumptive
\textquotedblleft rigorous limit\textquotedblright\ as $\omega\rightarrow0$.

Transverse fields do not possess a clear limit as the frequency approaches
zero. This is easily seen from the property, fundamental for propagating
fields, that there is a ponderomotive energy $U_{p}$ associated with such
fields that behaves as $U_{p}\sim\omega^{-2}$, making it energetically
impossible to reach zero frequency. Furthermore, the dipole approximation is
invalid when $U_{p}\gtrsim O\left(  mc^{2}\right)  ,$ so as $\omega
\rightarrow0$ this limit will be surpassed, the G\"{o}ppert-Mayer (GM) gauge
transformation\cite{gm} is not applicable, and tunneling concepts lose all
meaning for laser fields.

The above considerations establish that the tunneling model has limited
applicability for the description of atomic ionization, and that the Schwinger
limit applies only to longitudinal fields but not to laser fields.

\begin{figure}
[ptb]
\begin{center}
\includegraphics[
height=2.3488in,
width=3.0234in
]%
{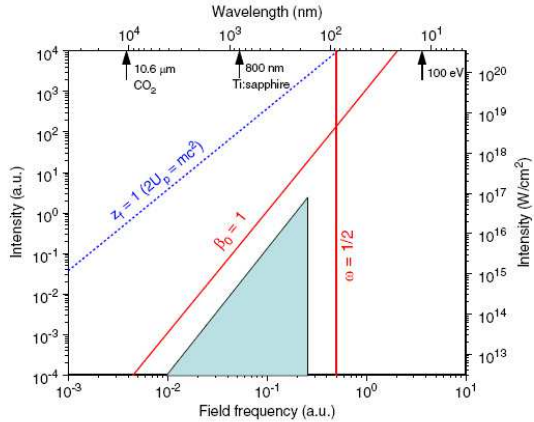}%
\caption{The region to the left of the line labeled \textquotedblleft%
$\beta_{0}=1$\textquotedblright\ is a region where magnetic forces become
strong enough to render invalid the dipole approximation. To the left of the
line $z_{f}=1$, conditions are relativistic, where dipole-approximation
results are entirely meaningless. These low-frequency limitations on the
dipole approximation are frequently overlooked. The figure is from Ref.
\cite{hr101}.}%
\label{1}%
\end{center}
\end{figure}

Whenever the magnetic field component of a laser field is strong enough that
$\mathbf{B}$ is important, and particularly if $U_{p}\gtrsim O\left(
mc^{2}\right)  $, then the GM gauge transformation is not valid for a
transverse field, and tunneling concepts do not apply to laser fields. These
limitations are illustrated graphically in Fig. \ref{1}, reproduced from Ref.
\cite{hr101}, that shows the onset of magnetic field effects as well as the
onset of a fully relativistic regime. Within the parameter space displayed in
Fig. \ref{1}, it is only the limited regime of frequencies and intensities
shown by the shaded area (the \textquotedblleft oasis\textquotedblright) where
tunneling concepts have any connection with laser phenomena (although
tunneling problems can arise even within the oasis).%

%TCIMACRO{\FRAME{ftbpFU}{3.0234in}{2.3488in}{0pt}{\Qcb{The region to the left
%of the line labeled \textquotedblleft$\beta_{0}=1$\textquotedblright\ is a
%region where magnetic forces become strong enough to render invalid the dipole
%approximation. To the left of the line $z_{f}=1$, conditions are relativistic,
%where dipole-approximation results are entirely meaningless. These
%low-frequency limitations on the dipole approximation are frequently
%overlooked. The figure is from Ref. \cite{hr101}.}}{\Qlb{1}}{fig1andp.eps}%
%width 3.0234in;  height 2.3488in;  depth 0pt;  original-width 5.8193in;
%original-height 4.5083in;  cropleft "0";  croptop "1";  cropright "1";
%cropbottom "0";  filename 'fig1andp.eps';file-properties "XNPEU";}} }%
%BeginExpansion
%EndExpansion

Figure \ref{1} illustrates that one must specify both the intensity $I$ and
the frequency $\omega$ of a laser field in order to establish the physical
conditions in which the ionization occurs. Alternatively, two dimensionless
intensity parameters may be specified. The notion that this two-dimensional
role can be played by a single parameter such as the Keldysh parameter
$\gamma_{K}$ is a seriously flawed concept\cite{hr76a,hr101,hr102,hr82}.

Transition rates for tunneling processes are characterized by the exponential
behavior%
\begin{equation}
\exp\left(  -C/E\right)  , \label{e}%
\end{equation}
where $C$ is a constant dependent on the properties of the system being
irradiated, and $E$ is the magnitude of the electric field. Attempts to show
the validity of tunneling behavior are often based on efforts to demonstrate
that the transition probability exhibits the behavior shown in Eq. (\ref{e}).

This goal is not to be confused with a quite different process. The process of
strong-field photon-multiphoton pair production from the vacuum, which was
first investigated in Ref. \cite{hr62a}, leads to a particular limit of a form
resembling tunneling behavior. The \textquotedblleft Toll-Wheeler
process\textquotedblright\cite{tollwh,hr62a} is one in which a constant
background electric field produces pairs upon interaction with a perturbative
QSE incoming field. The process is governed by a tunneling-like result
proportional to%
\begin{equation}
\exp\left(  -\frac{4}{3\chi}\right)  ,\quad\chi=\frac{\widetilde{\omega}}%
{m}\frac{E}{E_{crit}},\quad E_{crit}=\frac{m^{2}}{e}, \label{e1}%
\end{equation}
where natural units $\hbar=c=1$ are used, and $E_{crit}$ is the Schwinger
critical field. In the Toll-Wheeler process, the constant field replaces the
laser field of the photon-multiphoton pair-production process of Ref.
\cite{hr62a}. The tunneling exponential form of Eq. (\ref{e1}) is similar to
Eq. (\ref{e}), but $E$ is replaced by $\widetilde{\omega}E$, where
$\widetilde{\omega}$ is the frequency of the incoming field and $E$ is the
amplitude of the background field. The goal of the limiting process examined
in Ref. \cite{hr62a} is to show that the zero frequency limit for pair
production is the Toll-Wheeler result. From (\ref{e1}), this means that
$\widetilde{\omega}E\rightarrow0$ has to be examined. In contrast, Eq.
(\ref{e}), the usual tunneling limit, would be associated with large values of
the denominator quantity $E$. Not only are the limits of opposite type, but
the Toll-Wheeler parameter is a product of the frequency of one field with the
amplitude of the other. This is unrelated to a tunneling limit.

The 1964 Keldysh paper\cite{kel} on atomic ionization led to a focus on
tunneling as the mechanism of strong-field ionization. Keldysh employed the LG
in his work, where a QSE field is represented by scalar ($\phi$) and vector
($\mathbf{A}$) potentials $\phi^{LG}=-\mathbf{r\cdot E}\left(  t\right)
,\;\mathbf{A}^{LG}=0.$ The GM gauge transformation is employed to relate these
potentials to the VG potentials $\phi^{VG}=0,\;\mathbf{A}^{VG}=\mathbf{A}%
^{VG}\left(  t\right)  ,$where the VG is a Coulomb gauge in which the strong
form of the dipole approximation%
\begin{equation}
\mathbf{E}=\mathbf{E}\left(  t\right)  ,\mathbf{\quad B}=0 \label{c}%
\end{equation}
is employed. That is, the laser field is treated as if it were a QSE field.
This is explicit in the work of Perelomov, Popov, and Terent'ev
(PPT)\cite{ppt}. Nikishov and Ritus\cite{nr} independently generated a
tunneling method equivalent to that of PPT.

By the time the VG theory of atomic ionization was published\cite{hr80}, it
was regarded as a necessity to relate it to tunneling results. The Coulomb
gauge in dipole approximation is not the same thing as a QSE approach, which
is the reason that the theory of Ref. \cite{hr80} is not the same as a LG
theory. This is the cause of the differences in the two sets of results, both
of which are confusingly referred to as the Strong-Field Approximation (SFA).
This has led to the unjustified assumption that one theory should be exactly
gauge-equivalent to the other. Some of these differences were explored in Ref.
\cite{hr79}; they will be explicated further elsewhere. The GM gauge
transformation connects a QSE theory to a theory resembling the Coulomb gauge,
but where the strong-dipole-approximation conditions (\ref{c}) are applied.
These conditions are excessive, since they eliminate the propagation
properties of transverse fields. It is possible to apply a more limited
version of the dipole approximation that retains propagation properties and
hence the correct Maxwell equations for plane-wave fields, but this is not
gauge-equivalent to the LG.

The VG theory of Ref. \cite{hr80} has its provenance in a relativistic
formulation\cite{hr90,hrrev}, so its content is not the same as a LG theory.
It will be shown here that the closest correspondence to a \textquotedblleft
tunneling limit\textquotedblright\ predicted by the VG SFA does not yield a
tunneling result. This will be shown directly for circular polarization; for
linear polarization, reference will be made to already-published results.

The VG SFA ionization probability theory presented in Ref. \cite{hr80} will be
restated here with three changes from the original: (1) atomic units
($\hbar=m=\left\vert e\right\vert =1$) are used in place of \textquotedblleft
natural\textquotedblright\ units ($\hbar=c=1$); (2) the symmetrical
formulation of the transformation from configuration representation to
momentum representation%
\[
\phi_{i}\left(  \mathbf{p}\right)  =\frac{1}{\left(  2\pi\right)  ^{3/2}}\int
d^{3}\mathbf{r}\exp\left(  -i\mathbf{p\cdot r}\right)  \phi_{i}\left(
\mathbf{r}\right)
\]
is used in place of the asymmetrical form%
\[
\phi_{i}\left(  \mathbf{p}\right)  =\int d^{3}\mathbf{r}\exp\left(
-i\mathbf{p\cdot r}\right)  \phi_{i}\left(  \mathbf{r}\right)
\]
that was employed in \cite{hr80}; (3) angles are labeled by subscripts to
indicate whether they are momentum-space (subscript $\mathbf{p}$) or
configuration-space (subscript $\mathbf{r}$) angles.

For circular polarization, the differential transition probability for
ionization is%
\begin{align}
\frac{dW}{d\Omega_{p}}  &  =2\pi\sum_{n=n_{0}}^{\infty}p\left(  \frac{p^{2}%
}{2}+E_{B}\right)  ^{2}\left\vert \phi_{i}\left(  \mathbf{p}\right)
\right\vert ^{2}\left\vert J_{n}\left(  \varsigma_{c}\right)  \right\vert
^{2},\label{h}\\
n_{0}  &  =\left\lceil \left(  E_{B}+U_{p}\right)  /\omega\right\rceil
,\quad\varsigma_{c}=\alpha_{0}^{c}p_{\perp}=\alpha_{0}^{c}p\sin\theta
_{p},\quad\alpha_{0}^{c}=\left(  2z/\omega\right)  ^{1/2}, \label{i}%
\end{align}
where $\Omega$ is solid angle, $E_{B}$ is the initial binding energy (or
ionization potential), $\phi_{i}\left(  \mathbf{p}\right)  $ is the
momentum-space wave function of the field-free initial state, $\left\lceil
\;\right\rceil $ denotes the ceiling function (the smallest integer containing
the quantity within the brackets), $U_{p}$ is the ponderomotive energy of a
free electron in the field of frequency $\omega$, $\alpha_{0}^{c}$ is the
radius of the classical circular motion of an electron with ponderomotive
potential $U_{p}$ in the laser field of frequency $\omega$, $z$ is the
intensity parameter%
\begin{equation}
z\equiv U_{p}/\omega. \label{j}%
\end{equation}
and the axis of spherical coordinates is in the direction of propagation of
the laser field. A second, independent intensity parameter is%
\begin{equation}
z_{1}\equiv2U_{p}/E_{B}=1/\gamma_{K}^{2}, \label{m}%
\end{equation}
where $\gamma_{K}$ is the Keldysh parameter.

\begin{figure}
[ptb]
\begin{center}
\includegraphics[
height=2.3488in,
width=3.0407in
]%
{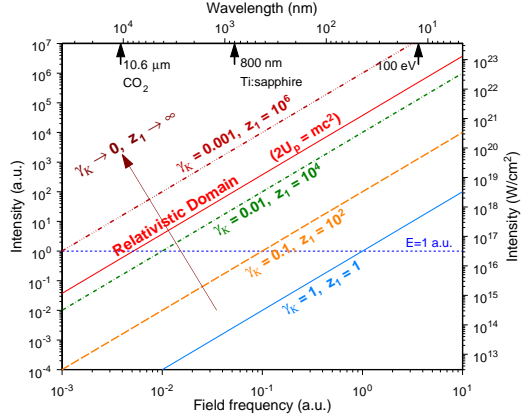}%
\caption{This figure shows that the Keldysh parameter $\gamma_{K}$ cannot
serve as a single scaling parameter since each line of constant $\gamma_{K}$
connects regions with completely different physical properties. It also shows
that as $\gamma_{K}\rightarrow0$, laser field conditions must become
relativistic, where the dipole approximation is invalid and a tunneling theory
cannot describe a laser field. The figure is adapted from Ref. \cite{hr82}.}%
\label{2}%
\end{center}
\end{figure}

Tunneling behavior would be associated with a field so intense that $z_{1}%
\gg1$ (or $\gamma_{K}\ll1$). It also requires that an innumerably large number
of photons are required to supply the ionization potential $E_{B},$ or
$E_{B}\gg\omega$. These conditions,
\begin{equation}
z_{1}\gg1,\quad E_{B}\gg\omega, \label{p}%
\end{equation}
can also be stated entirely in terms of the $z$ and $z_{1}$ intensity
parameters, as%
\begin{equation}
z_{1}\gg1,\quad2z\gg z_{1}. \label{n}%
\end{equation}
The nature of Fig. \ref{1} can be examined directly with respect to the
conditions (\ref{p}). Since $\omega$ is one of the coordinate axes, the second
of the conditions (\ref{p}) is just the vertical line at $\omega=0.5a.u.$,
with ground-state hydrogen used as an example. The condition $z_{1}\gg1$ (or
$\gamma_{K}\ll1$) is revealing when lines of constant $z_{1}$ (or constant
$\gamma_{K}$) are plotted\cite{hr82} as in Fig. \ref{2}. There it is seen that
one cannot employ very large values of $z_{1}$ (or small values of $\gamma
_{K}$) without getting into a domain that is unquestionably relativistic, and
hence where the dipole approximation is invalid. Actually, the domain marking
the failure of the dipole approximation ends where it is no longer possible to
ignore magnetic field effects, and it is the magnetic field that determines
the low-frequency boundary of the oasis region in Fig. \ref{1}.%

%TCIMACRO{\FRAME{ftbpFU}{3.0407in}{2.3488in}{0pt}{\Qcb{This figure shows that
%the Keldysh parameter $\gamma_{K}$ cannot serve as a single scaling parameter
%since each line of constant $\gamma_{K}$ connects regions with completely
%different physical properties. It also shows that as $\gamma_{K}\rightarrow0$,
%laser field conditions must become relativistic, where the dipole
%approximation is invalid and a tunneling theory cannot describe a laser field.
%The figure is adapted from Ref. \cite{hr82}.}}{\Qlb{2}}{fig2ntun.eps}%
%width 3.0407in;  height 2.3488in;  depth 0pt;  original-width 7.7236in;
%original-height 5.9525in;  cropleft "0";  croptop "1";  cropright "1";
%cropbottom "0";  filename 'fig2ntun.eps';file-properties "XNPEU";}} }%
%BeginExpansion
%EndExpansion

The literature on the tunneling limit largely ignores the low-frequency
constraints revealed in Figs. \ref{1} and \ref{2}, and examines only the
effects of $\gamma_{K}\ll1$, so that step will be taken here. It is a simple
matter to just employ an asymptotic form of the Bessel functions in Eq.
(\ref{h}) when the order $n$ is very large. The appropriate asymptotic form
depends on the relative magnitudes of the order $n$ and the argument
$\zeta_{c}$. From the definition of $\zeta_{c}$ in Eq. (\ref{i}) and the
symmetry of (\ref{h}), the angle $\theta_{p}$ can be confined to the first
quadrant where $\zeta_{c}$ is always positive. Starting with the inequality
$\left(  n-2z\right)  ^{2}\geq0,$ it follows that%
\begin{equation}
n^{2}\geq4nz-4z^{2}=4z\left(  n-z\right)  >0, \label{q}%
\end{equation}
where $n-z>0$ is a consequence of $n\geq n_{0}$, and $n_{0}$ given in
(\ref{i}) is rewritten as $n_{0}=\left\lceil z+\left(  E_{B}/\omega\right)
\right\rceil .$When $n_{0}\gg1,$ the ceiling function symbol in Eq. (\ref{i})
can be ignored, so that (\ref{i}) and (\ref{q}) give%
\begin{equation}
n\geq2z^{1/2}\left(  n-z\right)  ^{1/2}. \label{s}%
\end{equation}
The argument of the Bessel function is%
\begin{equation}
\zeta_{c}=\alpha_{0}^{c}p\sin\theta_{p}\leq\alpha_{0}^{c}p=\left(  \frac
{2z}{\omega}\right)  ^{1/2}p. \label{t}%
\end{equation}
The kinetic energy available to a photoelectron is $p^{2}/2=\left(
n-n_{0}\right)  \omega$, and energy conservation gives $p=\left(
2\omega\right)  ^{1/2}\left(  n-z-\epsilon_{B}\right)  ^{1/2}.$This gives an
upper limit on the argument of the Bessel function as%
\begin{equation}
\zeta_{c}\leq\left(  \frac{2z}{\omega}\right)  ^{1/2}\left(  2\omega\right)
^{1/2}\left(  n-z-\epsilon_{B}\right)  ^{1/2}=2z^{1/2}\left(  n-z-\epsilon
_{B}\right)  ^{1/2}<2z^{1/2}\left(  n-z\right)  ^{1/2}. \label{x}%
\end{equation}
Equations (\ref{s}) and (\ref{x}) establish that%
\begin{equation}
n>\zeta_{c}. \label{y}%
\end{equation}

Equation (\ref{y}) identifies the appropriate asymptotic Bessel function to be%
\begin{equation}
J_{n}\left(  x\right)  =J_{n}\left(  \frac{n}{\cosh\alpha}\right)
\approx\frac{\exp\left(  n\tanh\alpha-n\alpha\right)  }{\sqrt{2\pi
n\tanh\alpha}},\quad\cosh\alpha\equiv n/x, \label{z}%
\end{equation}
where the parameter $\alpha$ is such that%
\begin{align*}
\tanh\alpha &  =\frac{1}{n}\sqrt{n^{2}-x^{2}},\\
\alpha &  =\ln\left(  \frac{n}{x}+\frac{1}{x}\sqrt{n^{2}-x^{2}}\right)  ,\\
\exp\left(  -n\alpha\right)   &  =\frac{x^{n}}{\left(  n+\sqrt{n^{2}-x^{2}%
}\right)  ^{n}}.
\end{align*}
The squared asymptotic Bessel function is%
\begin{equation}
\left[  J_{n}\left(  x\right)  \right]  ^{2}\approx\frac{1}{2\pi\sqrt
{n^{2}-x^{2}}}\frac{x^{2n}\exp\left(  n^{2}-x^{2}\right)  }{\left(
n+\sqrt{n^{2}-x^{2}}\right)  ^{2n}}, \label{ab}%
\end{equation}
Field intensity dependence occurs within the parameter $x^{2}$ in Eq.
(\ref{ab}), where, from Eq. (\ref{t}), \ one has%
\begin{equation}
x^{2}=\zeta_{c}^{2}=\frac{2z}{\omega}p^{2}\sin^{2}\theta_{p}=4z\left(
n-z-E_{B}/\omega\right)  \sin^{2}\theta_{p}. \label{ac}%
\end{equation}
The parameter $z$ is proportional to $U_{p},$ and hence to field intensity
$I$, where $I$ follows from the square of the electric field strength.

The conclusion is that there nothing in the rate $dW/d\Omega_{p}$ in Eq.
(\ref{h}) that can exhibit tunneling dependence as in Eq. (\ref{e}). The VG
theory of Ref. \cite{hr80} is thus never of tunneling form for strong fields
of circular polarization.

This is not a surprising result in view of the fact that a tunneling theory
would predict that the photoelectron should emerge primarily in the direction
of the electric field, which is radial for circular polarization, The observed
direction\cite{bergues} as well as that predicted\cite{hr76a} in the VG SFA,
is azimuthal. That is, ionization caused by a circularly polarized laser field
is not even approximately a tunneling process.

For linear polarization, the differential transition rate is of a form
identical to Eq. (\ref{h}) apart from a replacement of the ordinary Bessel
function by the generalized Bessel function\cite{hr80,krs}%
\[
J_{n}\left(  \alpha_{0}^{l}p\cos\theta_{p},-\beta_{0}c\right)  ,
\]
where $\alpha_{0}^{l}$ is the amplitude of motion of the electron in the
direction parallel to the electric field when executing the well-known
\textquotedblleft figure-8\textquotedblright\ orbit, and $\beta_{0}$ is the
amplitude of motion parallel to the propagation vector. These amplitudes are%
\[
\alpha_{0}^{l}=2\left(  z/\omega\right)  ^{1/2},\quad\beta_{0}=z/2c.
\]
Using the tunneling conditions (\ref{p}), it was shown in Ref. \cite{hrvk1}
that the generalized Bessel function could indeed be placed in the tunneling
form, leading to an ionization rate proportional to%
\begin{equation}
\exp\left[  -\frac{2}{3}\frac{\left(  2E_{B}\right)  ^{3/2}}{E}\right]  .
\label{ad}%
\end{equation}
This exactly matches the form anticipated from Eq. (\ref{e}).

\begin{figure}
[ptb]
\begin{center}
\includegraphics[
height=2.3488in,
width=3.6236in
]%
{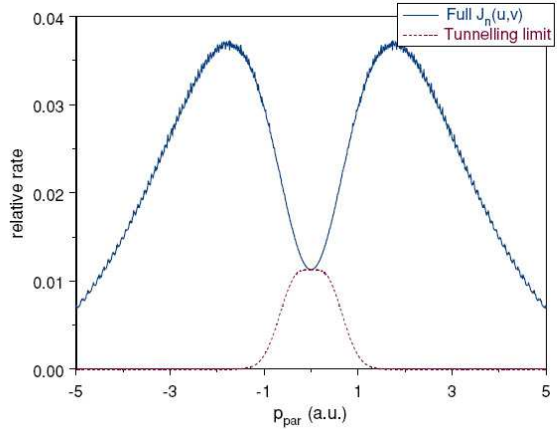}%
\caption{The momentum distribution in the ionization of ground-state neon as
calculated with a completely stated generalized Bessel function, and as
calculated with a tunneling approximation. Agreement occurs in the immediate
vicinity of $p_{par}=0$, but the failure of a tunneling approximation is
dramatically illustrated. The figure is from Ref. \cite{hrvk2}.}%
\label{3}%
\end{center}
\end{figure}

Despite the above result, it was shown in \cite{hrvk2} that only a portion of
the overall ionization rate possessed the form (\ref{ad}). Depending on the
particular ionization event considered, that part of the total rate showing
tunneling behavior was generally only a small portion of the complete rate,
and furthermore lacked many of the essential features of the total rate. An
example of this behavior is shown in Fig. \ref{3}, duplicating Fig. 2 of Ref.
\cite{hrvk2}, giving the momentum distribution of photoelectrons in the
ionization of ground-state neon.%

%TCIMACRO{\FRAME{ftbpFU}{3.6236in}{2.3488in}{0pt}{\Qcb{The momentum
%distribution in the ionization of ground-state neon as calculated with a
%completely stated generalized Bessel function, and as calculated with a
%tunneling approximation. Agreement occurs in the immediate vicinity of
%$p_{par}=0$, but the failure of a tunneling approximation is dramatically
%illustrated. The figure is from Ref. \cite{hrvk2}.}}{\Qlb{3}}{fig3ntun.eps}%
%width 3.6236in;  height 2.3488in;  depth 0pt;  original-width 7.2774in;
%original-height 4.6994in;  cropleft "0";  croptop "1";  cropright "1";
%cropbottom "0";  filename 'fig3ntun.eps';file-properties "XNPEU";}} }%
%BeginExpansion
%EndExpansion

An earlier attempt to exhibit tunneling behavior in a VG theory was done in
Section V of Ref. \cite{hr80}. It was shown there that it is possible to
expand a linear polarization ionization rate as a power series in the
parameter $\beta=\left(  n-n_{0}\right)  /n_{0}$. The zero-order term has the
form of a tunneling exponential. However, as the field intensity increases,
the value of $\beta$ increases, so that the zero-order term is no longer
representative of the actual rate.

\textit{The final conclusions reached here are twofold:}

\begin{enumerate}
\item \textit{The so-called tunneling limit for laser-induced ionization has
no relevance since such a limit does not exist for transverse electromagnetic
fields, as shown graphically in Figs. \ref{1} and \ref{2}. A tunneling limit
does exist for longitudinal fields, but laser fields are transverse;
fundamentally a different species of electromagnetic phenomena.}

\item \textit{Attempts to exhibit the algebraic behavior of tunneling rates of
the type of Eq. (\ref{e}) from velocity-gauge or Coulomb-gauge calculations
have led to outright failure (as for circular polarization) or to
unrepresentative segments of the complete rate.}
\end{enumerate}


\begin{thebibliography}{99}                                                                                               %


\bibitem {sauter}F. Sauter, Z. Phys. \textbf{69}, 742 (1931).

\bibitem {schwinger}J. Schwinger, Phys. Rev. \textbf{82}, 664 (1951).

\bibitem {kel}L. V. Keldysh, Zh. Exp. Teor. Phys. \textbf{47}, 1945 (1964)
[Sov. Phys. JETP \textbf{20}, 1307 (1965)].

\bibitem {gm}M. G\"{o}ppert-Mayer, Ann. Phys. (Leipzig) \textbf{9}, 273 (1931).

\bibitem {hr101}H. R. Reiss, Phys. Rev. Lett. \textbf{101}, 043002 (2008);
\textbf{101}, 159901(E) (2008).

\bibitem {hr76a}H. R. Reiss, Phys. Rev. A \textbf{76}, 033404 (2007).

\bibitem {hr102}H. R. Reiss, Phys. Rev. Lett. \textbf{102}, 143003 (2009).

\bibitem {hr82}H. R. Reiss, Phys. Rev. A \textbf{82}, 023418 (2010).

\bibitem {hr62a}H. R. Reiss, J. Math. Phys. \textbf{3}, 59 (1962).

\bibitem {tollwh}J. S. Toll and J. A. Wheeler, private communication (1958).

\bibitem {ppt}A. M. Perelomov, V. S. Popov, and M. V. Terent'ev, Zh. Exp.
Teor. Phys. \textbf{50}, 924 (1966) [Sov. Phys. JETP \textbf{23}, 924 (1966)].

\bibitem {nr}A. I. Nikishov and V. I. Ritus, Zh. Exp. Teor. Phys. \textbf{50},
255 (1966) [Sov. Phys. JETP \textbf{23}, 168 (1966)].

\bibitem {hr80}H. R. Reiss, Phys. Rev. A \textbf{22}, 1786 (1980).

\bibitem {hr79}H. R. Reiss, Phys. Rev. A \textbf{19}, 1140 (1979).

\bibitem {hr90}H. R. Reiss, Phys. Rev. A \textbf{42}, 1476 (1990).

\bibitem {hrrev}H. R. Reiss, Prog. Quant. Elec. \textbf{16}, 1 (1992).

\bibitem {bergues}B. Bergues, Y. Ni, H. Helm, and I. Yu. Kiyan, Phys. Rev.
Lett. \textbf{95}, 263002 (2005).

\bibitem {krs}V. P. Krainov, H. R. Reiss, and B. M. Smirnov, \textit{Radiative
Processes in Atomic Physics }(Wiley, New York, 1997) Appendix J.

\bibitem {hrvk1}H. R. Reiss and V. P. Krainov, J. Phys. A \textbf{36}, 5575 (2003).

\bibitem {hrvk2}H. R. Reiss and V. P. Krainov, J. Phys. A \textbf{38}, 527 (2005).
\end{thebibliography}
\end{document}